\documentclass[12pt]{article}
\advance\textheight by \topskip
\newcommand{\be}{\begin{equation}}
\newcommand{\ee}{\end{equation}}
\def\bea{\begin{eqnarray}}
\def\eea{\end{eqnarray}}
\def\der{\partial}

\newcommand{\ti}{\tilde}

\newcommand{\ft}[2]{{\textstyle\frac{#1}{#2}}}

\usepackage{graphics}

\begin{document}

 %%%%%%%%%%%%%%%%%%%%%%%%%%%%%%%%%%%%%%%%%%%%%%%%%%%%%%%%%%%
\begin{titlepage}
\begin{flushright}
SU-ITP 00/27\\
hep-th/0010271
\end{flushright}
\vspace{.5cm}
\begin{center}
\baselineskip=16pt {\LARGE \bf
Excision of Singularities \\

by Stringy Domain Walls

}
\vskip 1.5 cm
{\large  Renata Kallosh${}^a$,
Thomas Mohaupt${}^{a,b}$  and Marina Shmakova${}^c$} \\
\vskip 1 cm
{\small 
${}^a$ Physics Department, Stanford University, Stanford, CA 94305 \\
${}^b$ Fachgruppe Theoretische Physik, Universit\"at Halle, D-06099 Halle \\
${}^c$ California Institute for Physics and Astrophysics \\
366 Cambridge Avenue Palo Alto, CA 94306\\

\vskip 0.6cm
}
\end{center}
\vskip 2 cm
\begin{center}
{\bf Abstract}
\end{center}
{\small We study  supersymmetric
domain walls on ${S_1/{\bf Z}_2}$ orbifolds. The supergravity solutions in the bulk
are given by the attractor equation associated with Calabi-Yau spaces and have
a naked space-time singularity at some $|y_s|$.
We are looking for  possibilities
to cut off this singularity with the second wall by a stringy mechanism.
We use the collapse of the CY cycle at $|y_c|$ which happens before and at a finite
distance from the space-time singularity. In our example with three K\"ahler moduli
the second wall is at the end of the moduli space at $|y_c|$ where also the 
enhancement of $SU(2)$ gauge symmetry takes place so that $|y_e| = |y_c| < |y_s|$.
The physics of the excision of a naked singularity 
via the enhan\c{c}on in the context of domain wall has an interpretation on the heterotic side
related to $R\rightarrow1/R$ duality. The position of the 
enhan\c{c}on is given by the equation $R(|y_e|)=1$.

}
\end{titlepage}

The supersymmetric domain wall solutions of D=5 N=2 $U(1)$
gauged supergravity \cite{GST} with brane sources on 
$S^1/{\bf Z}_2$ orbifolds have been described recently
in \cite{BerKalVPro:2000}.  It has been observed there that 
in the context of Calabi-Yau (CY) compactifications
the collapse of CY cycles 
may put some restrictions on the
distance between the walls.\footnote{The study of the collapsing
CY cycles was reported recently in \cite{greene}.} 
In  this paper we will study this type of domain
walls both for D=5 N=2 $U(1)$
gauged supergravity \cite{GST} (GST model) and for
Calabi-Yau compactifications
of 11D supergravity with fluxes turned on. The later is
the five-dimensional heterotic M-theory \cite{Wit:1996,LukOvrSteWal:1998}
obtained by a reduction on a CY threefold of 
Horava-Witten M-theory \cite{HoravaWitten} on $S^1/{\bf Z}_2$ (HW model).
The explicit form of the solution with general dependence on the
vector multiplets is obtained for both models by solving 
the generalised attractor equation
\cite{FKS,critical,sabra,GaiMahMohSab:1998}. Since the domain wall solutions 
\cite{BerKalVPro:2000,LukOvrSteWal:1998}
of the two models behave very similar, we will discuss them in parallel.

The purpose of this paper is to find a possibility to remove the space-time
singularity of the domain wall solution via some particular property of the CY
space. Specifically we would like to find a situation when the collapse
of the CY cycle at $| y_{c}|$ happens closer to the first wall which is at
$y=0$ and at a finite distance from the space-time
singularity $|y_s|$, so that
\be
| y_c|<|y_s| \;.
\label{finite}\ee
In the case of excision of repulson singularities by the 
enhan\c{c}on mechanism \cite{Johnson:2000qt}, the distance between the repulson and enhan\c{c}on   is finite. The hope therefore is that also for some domain walls the analogous situation may be possible, particularly if enhancement of gauge symmetry is somehow involved.
The finite distance between the naked singularity of the supergravity solution
and the position of the collapse 
of the CY cycle  may allow us 
to use the physics of string theory already at the end of the moduli space which in this case excludes the singularity of the general relativity as  unphysical. The generic  interest to such mechanism is supported by some interesting recent investigations of the brane world scenarios \cite{naked} where the naked singularities may be present in solutions of the Einstein equations and need to be removed by a natural stringy type mechanism. 

Our walls are supersymmetric everywhere, including the position of the branes \cite{BerKalVPro:2000}, therefore they do not directly address the problems of models \cite{naked}.  But due to supersymmetry in our model 
 the matching condition 
for the solutions are satisfied automatically on both walls. We have more control over the situation and may clearly indicate conditions when a natural mechanism of stringy excision of singularities is available.

We  have found  
that in most cases the singularities of the CY space tend to
coincide with the space-time singularity of the domain wall solutions, i.e.
\be
|y_c|=|y_s| \;.
\ee
Only in some special cases we will find the singularities in 
space-time  and CY space at some finite
distance in $y$-direction from each other as in eq. (\ref{finite}). 
Within the 
classification of the possible behaviour of the CY manifold at the
boundary of the K\"ahler cone \cite{critical,witten} we consider the
special case when a complex divisor D is collapsing to a curve E 
of $A_1$ singularities, so that there
is an $SU(2)$ enhancement of gauge symmetry at the boundary.
The position $|y_c|$ where the cycle collapses therefore corresponds to the  
position $|y_e|$ of the enhan\c{c}on. We will find examples where
\be
| y_c|=|y_e|<|y_s| \;.
\label{ourexamples}
\ee
Domain wall solutions of the two models have a metric of the form
\cite{BehGuk:2000}
\be
ds^2 = a^2(y) dx^\mu dx^\nu \eta_{\mu\nu} + a^{2 \gamma}(y) dy^2 \;,
\ee
where $\gamma = -2$ is the GST model and $\gamma = 4$ is the
HW model.\footnote{We use the notation and conventions of
\cite{BerKalVPro:2000}.}
In both cases
the function $a(y)$ and the scalar fields
are determined in terms of harmonic functions through
generalized  attractor  equations 
\cite{critical,sabra} which require that
the rescaled variables $\ti{h}^I \equiv a(y) h^I$ have to
satisfy
\be
C_{IJK} \tilde{h}^J \tilde{h}^k = H_I(y)\ ,
\label{GenStabEq1}
\ee
where $H_I(y)$ are harmonic functions. 
Then the physical scalars
are given by either solving the
hypersurface constraint or by using the
the ratios $\ti{h}^x/\ti{h}^0$ 
and the metric
is determined by
\be
a^3(y)  = C_{IJK} \ti{h}^I \ti{h}^J \ti{h}^K = \ti{h}^I H_I \;.
\ee
In the HW model one additional scalar enters non-trivially into the solution.
This scalar is the overall volume $V$, or `breathing
mode' of the Calabi-Yau space. Since there is no solution with constant
$V$,  there are no AdS vacua in the HW case 
in contrast to domain walls in 5d
supergravity \cite{BerKalVPro:2000}.  For our purpose it is important that $V$
is uniquely
determined by the vector multiplet scalars. In fact $V$ is just
some power of $a$, and therefore a rational function of the
harmonic functions:
\be
V = a^6 = (C_{IJK} \ti{h}^I\ti{h}^J\ti{h}^K)^2 \ .
\ee
As a consequence the flow through moduli space is the same as in
the GST model.
The two models differ in the precise form of the
space-time metric and by the fact that $a^6$ in the HW case is the volume
of the internal space.

Following \cite{BerKalVPro:2000} we consider a setup where
the 5-th direction is a $S^1 / {\bf Z}_2$ orbifold with 3-branes at the
fixed points $y=0$ and $y=\ti{y}$, which act as sources for the harmonic
functions:
\bea
\der^2_y H_I& =& - 2 g{q}_I [ \delta(y) - \delta(y-\tilde{y})]\nonumber \\
H_I &=& c_I - 2 g{q}_I |y| \, .
\label{harm}\eea
Concerning space-time singularities both models behave very similarly.
Components of curvature tensors become singular if either
$a=0$ or if its derivatives diverge.
The corresponding Ricci scalar is
\be
{\cal R} =- 4 a^{ -2 - 2 \gamma} \left(
(3 - 2 \gamma) (a')^2 + 2 a a'' \right) \ .
\label{RicciScalar}
\ee
To find explicit domain wall solutions we  consider some CY spaces
with 3 K\"ahler moduli
\cite{Candelas,LSTY} for which
the
relevant prepotential was identified in 5d supergravity and the attractor
equations have been solved. Many of such solutions are displayed and analysed
in \cite{critical,GaiMahMohSab:1998} for the extended
K\"ahler cone of a CY which is an elliptic fibration
over the Hirzebruch surface ${\cal F}_1$.  The extended K\"ahler cone
consists of two K\"ahler cones related by a flop 
transition.\footnote{Recently an extensive
study of the flop transition in the context of supersymmetric domain walls was
performed in \cite{GSS}.} We will refer to the two CY compactifications
as model III and model II, respectively. 
Model I forms a particular boundary of the moduli
space of the model II. The moduli space of model III has a boundary
where $SU(2)$ enhancement occurs in the way described above. Moreover
the metric on the moduli space is finite at this boundary.
As explained in \cite{Candelas,LSTY} the region III
CY is related to two other CY spaces by deformation of the base of the
elliptic fibration into the Hirzebruch surfaces
${\cal F}_0$ and ${\cal F}_2$, respectively. These models likewise
have a boundary with $SU(2)$ enhancement, and the physics close to
the boundary is completely isomorphic to the one of the region III
${\cal F}_1$ model. Though we will discuss the region III model
for definiteness, our results will be automatically valid for these
models as well. The M-theory compactifications on the elliptically 
fibered CY spaces with bases ${\cal F}_0, {\cal F}_1, {\cal F}_2$ 
have a dual description by compactifications of the $E_8 \times E_8$
heterotic string on $K3 \times S^1$ with instanton numbers
$(12,12), (13,11), (14,10)$, respectively. To be precise this 
duality is known to be valid in absence of $G$-flux inside the
M-theory CY. We will later use the heterotic picture to describe
the $SU(2)$ enhancement in a simple way, assuming that the duality
is still valid in presence of $G$-flux. Since switching on $G$-flux
does not interfere with the mechanism underlying gauge symmetry enhancement
this is a reasonable assumption.

Let us return to the M-theory compactification on the CY with 
base ${\cal F}_1$. 
The boundaries  of the extended  K\"ahler cone are defined by the
collapse of some cycles to zero volume.  The whole
picture is shown in Fig. 2 of \cite{critical}. Some of the boundaries have a
vanishing metric of the moduli space, some have an infinite metric. Equations
of motion relate the space-time curvature ${\cal R}$ with expressions 
which depend on
moduli space metric, 
$g_{xy}(\phi^x)' (\phi^y)' a^{-2\gamma}$, as well as
with expressions depending 
on the inverse moduli space metric, $W_{,x}g^{xy}W_{,y}$.
This indicates that it is likely that the space-time curvature is
infinite
if the moduli space metric $g_{xy}$ or its inverse $g^{xy}$ are
infinite.\footnote{The space-time curvature may still
be finite if  $(\phi^x)' ,
 a^{-2\gamma}, W_{,x}$ vanish at the singularity of the moduli space.}
We have studied  several cases  explicitly and found that they indeed
have coinciding singularities of the space-time and the moduli space.

We proceed therefore directly with the domain walls of the model III,
 which has a boundary with finite metric.  The
classical prepotential for this model was derived in \cite{LSTY}.
In terms of the variables adapted to the K\"ahler cone,
the prepotential is
\be
{\cal V} = \ft43 (h^1)^3 + \ft32 (h^1)^2 h^2
+ \ft12 h^1 (h^2)^2 + (h^1)^2 h^3 + h^1 h^2 h^3 = 1
\ee
and the K\"ahler cone is simply $h^I > 0$. In the
new variables 
\bea
h^1= U  \;,\qquad 
h^2 =T - \ft12 U - W \;,\qquad
h^3 = W - U 
\eea
the prepotential becomes
\be
{\cal V} = \ft5{24} U^3 + \ft12 U T^2 - \ft12 U W^2 + \ft12 U^2 W  = 1 \;.
\label{prepo}
\ee
The K\"ahler cone is:
$
W > U > 0 \mbox{   and   }
T > W + \ft12 U
$.
There are 3 boundaries when either of $h^I$ vanishes: i) $U=0 \Leftrightarrow
h^1= 0$: here the metric
of moduli space becomes singular, ii) $T = W + \ft{U}2 \Leftrightarrow h^2 =0$:
the metric of moduli space is regular and one has non-abelian gauge symmetry
enhancement, iii) $W=U
\Leftrightarrow h^3=0$: there is a 
flop transition, and again the metric is regular. 
We can solve (\ref{prepo}) for $T$:
$
T = \sqrt{ \ft2{U} \left(
1 - \ft{ 5U^3}{24} - \ft{U^2 W}{2} + \ft{ U W^2}{2}  \right) }
$
and keep as independent
scalars
$
\phi^x = (U,W).
$
By looking at the resulting moduli space metric $g_{xy}$ with determinant
$
\det g_{xy} \simeq \frac{ 12 ( 3 - 4 U^3 )}{24 U - 5 U^4 - 12 U^3 W + 12 U^2 W^2}
$
we recover the picture given in Figure 2 in \cite{critical}:
$U$ varies within a finite interval, whereas $W$ varies from
$0$ to $\infty$ at $U = 0$ and is cut-off by the curves
$U=W$ and $T-W - \ft12 U $ for positive $W$:
\be
0 < U < (\ft34)^{1/3} \;,\qquad
U < W < \frac{3 - U^3}{3 U^2} \;.
\ee
The stabilization equations (\ref{GenStabEq1}) are a 
system of quadratic equations
for the rescaled scalars $\ti{h}^I = a(y) h^I =: \ti{U}, \ti{T}, \ti{W}$.
For our model they are solved
following \cite{critical} by\footnote{The harmonic functions $H_I= c_I - d_I|y|$ used here and below are different  from the harmonic functions in  \cite{BerKalVPro:2000} and shown in eq. (\ref{harm}) by a factor of 6. The first term $c_I$ is arbitrary, so we will call it again $c_I$, the second term is $d_I=1/3 gq_I$. 
 This normalization of the
harmonic functions differs by a factor of 2 from the one
used in \cite{GaiMahMohSab:1998} and more recently 
in \cite{GSS}. When comparing to the solution of the attractor equations
given in \cite{critical} one should also rescale the charges.
}
\bea
\ti{U}= \sqrt{ \alpha - \sqrt{\alpha^2 - \beta}} \;, \qquad 
\ti{T} = \frac{H_T}{ \ti{U}} \;, 
\qquad 
\ti{W} = \ft12 \ti{U}  -
\frac{H_W}{ 2 \ti{U}}  \;,
\label{SolutionIII-WTU}
\eea
where
\be
\alpha = \ft14 (H_U + \ft12 H_W) \;, \qquad
\beta = \ft18 ( H_T^2 - H_W^2) \;.
\ee
One needs to impose that the scalars are real and inside the
K\"ahler cone.  Therefore the harmonic function are subject to
the inequalities
\be
\ft23 H_U \geq H_T \geq - H_W \geq \ft29 \left(
H_U - \sqrt{H_U^2 - \ft94 H_T^2} \right) \;,
\label{KC-harmonic}
\ee
which are mutually consistent. The boundary
$T = W + \ft{U}2$ corresponds to
$2 H_U = 3 H_T$, whereas the boundary $W=U$ corresponds to
$-H_W = \ft29 ( H_U - \sqrt{H_U^2 - \ft94 H_T^2})$ and the
boundary $U=0$ corresponds to $H_T = - H_W$. We would like
to mention that the second branch of the attractor equations
found in \cite{GaiMahMohSab:1998} does not describe a solution
inside the K\"ahler cone, as can be verified by a full analysis
of the constraints. 

Let us show that for
generic
values of the parameters of the harmonic functions the collapse of the
modulus $h^2$ is taking place at the point $| y_c|$ which is at the finite
distance from the space-time singularity.

First of all we have to find out under which conditions
space-time curvature can diverge. Looking at the formula 
(\ref{RicciScalar}) for the Ricci scalar we find that this happens if 
either $a=0$ or one of its derivatives diverges.\footnote{
The divergences in $a'$ cancel
(for the generic case where $a\not=0$)
as required by the relation $a \dot{a} \sim W$ implied by very 
special geometry, see equation (6.19) in \cite{BerKalVPro:2000}.  
However $a''$ is generically singular when $U=0$.}  
The same is
true for the components of the Ricci tensor and of the
Riemann tensor, which we did not display explicitly. 
The only point within the extended K\"ahler cone where
$a$ vanishes is $U=W=0, T=\infty$. At this point the moduli
space metric is infinite.
Divergences in the derivatives
of $a$ occur when either $\alpha = \sqrt{\alpha^2 - \beta}$ or
$\alpha^2 = \beta$. The first case corresponds to $U=0$, which 
is a boundary of the K\"ahler cone on which the moduli space metric
diverges. This includes the point where $a=0$ as a subcase.
Thus on the  boundary $U=0$ one finds the expected coincidence of 
space-time singularities
with moduli space singularities.
The only kind of space-time singularities which need
to concern us here are the ones related to $\alpha^2 = \beta$.

The equation $\alpha^2 = \beta$ has no solutions if 
$9 H_T^2 < 4 H_U^2$ which corresponds to $T > W + \ft12 U$.
Therefore no space-time singularity can occur as long
as the moduli are inside the K\"ahler cone. If $9 H_T^2 > 4 H_U^2$
then $\alpha^2 = \beta$ has two solutions,
$H_W = - \ft29 ( H_U \pm \sqrt{8} \sqrt{\ft94 H_T^2 - H_U^2})$.
Thus the generic situation is that one {\em first} crosses the 
enhancement boundary $T = W + \ft12 U$ and {\em then} runs into a 
space-time singularity at a finite distance.
If $9 H_T^2 = 4 H_U^2$, which is precisely true on 
the enhancement boundary, then $\alpha^2 = \beta$ has one solution
given by $2H_U + 9 H_W=0$. Thus
the only possibility for the space-time singularity 
to coincide with the boundary of moduli space is when the parameters
are fine tuned such that $2 H_U(y_c) = 3 H_T(y_c) = - 9 H_W(y_c)$.
The corresponding point in moduli space is the intersection point
of the enhancement boundary $T = W + \ft12 U$ with the flop boundary
$U=W$. At this point the metric on moduli space is degenerate, which 
nicely fits with our observation that a singularity in moduli space
generically induces a singularity in space-time.

The coordinates $W,T,U$ cover both regions of model III and II and allow
to analyse the flop transition in the framework of special geometry, as shown
in \cite{critical}. It is also interesting to use the description of the region
 III in $STU$ parametrization. In this parametrization we can  also 
 show that the singularities of the space-time and CY space are at finite
distance and we will give a numerical example. Moreover, the interpretation of the enhanc\c{o}n type physics
 in terms of $T$-duality is
manifest.

After the substitution 
$W = S' - \ft12(T' - U')$, 
$T = S' + \ft12 T'$,
$U = U'$, 
the region III prepotential takes the form
\be
{\cal V} = S'T'U' + \ft13 U'^3 \;.
\label{RegionIII-STU}
\ee
The original CY K\"ahler moduli are now related to the heterotic string variables as
follows:
\be
h^1=U' \ , \qquad h^2= T'-U'\ ,  \qquad h^3 = S' -{1\over 2} (T'+U')\ .
\ee
The solution in these variables is \cite{gaida}
\bea
\ti{U}' = \ft12 \sqrt{ H_U' - \sqrt{ (H_U')^2 - 4 H_S' H_T'}} \;,\qquad
\ti{T}' = \frac{H_S'}{2 \tilde{U}'} \;, \qquad 
\ti{S}' = \frac{H_T'}{2 \it{U}'} \;.
\label{SolutionIII-STU}
\eea
Note that the harmonic functions are now associated with the
primed variables.
The boundaries of region III are
\bea
U>0 &\Leftrightarrow&  U' > 0 \;,\nonumber \\
W>U & \Leftrightarrow& S' > \ft12 ( T' + U')  \;,\nonumber \\
T> W + \ft{U}2 & \Leftrightarrow &  T' > U' \;.
\label{Boundaries-STU}
\eea
For convenience we drop the primes on moduli and harmonic functions
in the rest of the paper, denoting moduli simply by $S,T,U$. One should
keep in mind that the $T$-variables in both parametrizations are
different!

Let us look at the moduli space metric.
We solve the hypersurface equation
$
{\cal V} = STU + \ft13 U^3 = 1
$
for $S$:
$
S = \frac{3 - U^3}{ 3 TU}
$.
The determinant of the vector kinetic matrix is
$
\det G_{IJ} \simeq 1 - \ft43 U^3 \;.
$
Thus $0 \leq U \leq (\ft34)^{1/3}$ as expected, because the
$U$ variable is the same in the $TUW$ and $STU$ parametrization.
The determinant of the scalar kinetic term is
\be
\det g_{xy} \simeq \frac{3 - 4 U^3}{ T^2 U^2} \;,
\ee
implying $0 < U < (\ft34)^{1/3}$ and $T \not=0$.
Since $T$ is positive for our CY moduli space, the
moduli space metric is regular for
\be
0 < U < (\ft34)^{1/3} \mbox{    and    }
0 < T \;.
\ee
In particular it becomes singular on $U=0$, which is
a boundary from the CY point of view (tensionless strings). On the boundary
$U=T$ (symmetry enhancement) it is regular,
as long as $U$ takes allowed values.
The third boundary (flop) is given by
\be
S(T,U) = \ft12 ( T + U) \;,
\ee
which can be solved for $T$ as
$
T(U) = - \ft12 U + \ft{\sqrt{3}}{6 U} \sqrt{24 U - 5 U^4}
$.
Note that $T(U)$ is positive for all $0<U< (\ft34)^{1/3}$.
Therefore the moduli space metric is regular along the
flop line.
The reality of the 'inner' and 'outer' roots in (\ref{SolutionIII-STU})
imposes 
\be
H_U^2 > 4 H_S H_T  > 0 \;.
\label{ConstraintIIISTU1}
\ee
A further look on (\ref{SolutionIII-STU}) and 
(\ref{Boundaries-STU}) tells us that the harmonic function 
$H_S, H_T, H_U$ have to be positive. 
When combining this with (\ref{ConstraintIIISTU1})
then all expressions are real and $U >0$.
The other boundaries are $T \geq U$ and $S \geq \ft12 (T+U)$.
The condition $T \geq U$ takes a very simple form,
\be
H_U  \geq H_S + H_T \;,
\label{ConstraintIIISTU2}
\ee
which is compatible with (\ref{ConstraintIIISTU1}).
We are interested in the limit $H_U \rightarrow H_S + H_T$.
We still 
have to implement the constraint that $h^3$ is positive, 
which in these variables is:
$S >  \ft12 (T+U)$.  We will impose the stronger constraint $S>T$ which yields
a simpler 
constraint on the harmonic functions and has the additional advantage to 
guarantee that
our solution is also inside the K\"ahler cones of the related ${\cal F}_0$ and
${\cal F}_2$ models.  For these models the prepotential likewise
can be brought to the form (\ref{RegionIII-STU}). However the
boundaries of the K\"ahler cones are different. For the
${\cal F}_2$ model the K\"ahler cone is defined by 
$S > T > U > 0$, whereas for the ${\cal F}_0$ model one
has $S,T > U > 0$. Note that all models share the $U=T$ boundary,
though the other boundaries are different. Moreover when imposing
the strongest constraint $S > T > U > 0$ we can discuss the limit
$T-U \rightarrow 0$ for all three models simultaneously. 
Now $S > T$ simply
implies
\be
H_T > H_S
\label{ConstraintIIISTU3} \;.
\ee
The constraints we found are compatible: evaluating
(\ref{ConstraintIIISTU1}), when (\ref{ConstraintIIISTU3})
is saturated, gives $(H_T - H_S)^2 \geq 0$.
Thus the boundary $U=T$ requires that $H_U = H_S + H_T$. This defines its position as
\be
|y_c| = \frac{c_U - c_T - c_S}{ d_U - d_T - d_S} \;.
\ee
%As in the previous case, the only possibility to have divergent derivatives of the moduli at $|y_c|$ is by satisfying the condition that 
%\be
%H_U^2 = 4 H_S H_T  = (H_S + H_T)^2 \quad \Rightarrow \quad  (H_T - H_S)^2 = 0\;.
%\label{third'}
%\ee
%\be
%|y_c| = \frac{ c_T - c_S}{ d_T - d_S}
%\ee
%Thus only when $\frac{c_U - c_T - c_S}{ d_U - d_T - d_S}= \frac{ c_T - c_S}{ d_T - d_S}$ one can find the space-time singularity at the boundary. In generic case this is not satisfied and the singularity of space time is outside.

A closer inspection of the analytic form of the Ricci scalar ${\cal R}$
and of the function $a$ and its space-time derivatives shows that
curvature singularity precisely occurs when $H_U^2 = 4 H_S H_T$.
Given the inequalities (\ref{ConstraintIIISTU1}-\ref{ConstraintIIISTU3}) 
we see that this can never happen
inside the K\"ahler cone. Moreover the generic situation is that
the space-time singularity is encountered after crossing the $T=U$
boundary. The only possibility to have the space-time singularity
coincide with the boundary of moduli space is to fine tune the 
parameters such that $H_T(y_c) = H_S(y_c)$ coincides with
$H_U^2(y_s) = 4 H_T(y_s) H_U(y_s)$ at $y=y_s=y_c$. At such a point one
has $S=T=U$ or $H_U^2(y_c) = 4 H_S^2(y_c) = 4 H_T^2(y_c)$. In terms
of the parameters in the harmonic function this means that one
must arrange 
$\frac{c_U - c_T - c_S}{ d_U - d_T - d_S}= \frac{ c_T - c_S}{ d_T - d_S}\; \Leftrightarrow \; |y_c|= |y_s|$. Generically this condition is not satisfied and therefore $|y_c|<|y_s|$.

Now we  can set up an example of a solution running into the
enhancement boundary. We take care of the constraint $H_T > H_S$ by setting
$H_T = 2 H_S$. It will turn out that this will lead to relatively simple
analytic expressions.
Note that this choice implies that at the enhancement 
boundary $U=T$
one has $S=2U$ or in terms of $h^I$: at $h^2=0$ one has $h^1=h^3$, see Fig. 2. 
%This choice simplifies the solution. Moreover we have
%$
%H_U \geq 3 H_S \mbox{   and    }
%H_U^2 > 8 H_S \ .
%$
%The first inequality corresponds to $T \geq U$, and 
%by construction it is saturated before the second one. 
%Violating the second inequality make the solution complex and
%at the point where it is saturated one encounters a naked
%space-time singularity, as we will see below. 

The harmonic functions take the form
$
H_I = c_I - d_I |y|\ 
$
dictated by the presence of two space-time boundaries. The constants
$c_I$ define the initial condition on the first space-time boundary whereas
the slopes $d_I$ determine how the solution flows through moduli space.
The $c_I$ are undetermined integration constants, which are only
restricted by the fact that all scalars should be inside the K\"ahler
cone at $y=0$ and by the conventional normalization $a=1$ that we impose
on the metric at $y=0$. On the other hand the $d_I$ are, in the context of a Calabi-Yau
compactification with flux, determined by the sources of flux put on the
boundaries,  \cite{BehGuk:2000,GSS}. We will choose some values for 
$d_I$ to simplify the calculations  and not try to connect these values to particular sources of fluxes, since we have shown that the picture is
generic. It will not change when taking different slopes, as long
as the solution runs into the enhancement boundary without reaching any other
boundary of moduli space first.

Now we choose initial data. We have $c_T = 2 c_S$ and have
to impose $c_U > c_T + c_S = 3 c_S$. For definiteness we
take $c_U = 4 c_S$. Then $c_S$ is fixed by the normalization
condition $a(0) = 1$. This can be solved exactly with the
result
\be
c_S = \left( \ft{45}{49} + \ft{9}{49 \sqrt{2}}
\right)^{1/3} \;.
\ee
Now we have to set the slope. We already decided to take
$d_T = 2d_S$. Then the boundary $T=U$ is reached once
the inequality
\be
|y| \leq |y_e| = \frac{c_U - c_T - c_S}{ d_U - d_T - d_S}
= \frac{c_S}{d_U - 3 d_S}
\ee
is saturated. We are free to choose $d_U > 3 d_S$. For
definiteness we take $d_U = 10$ and $d_S =1$.
The analytical value of $|y_e|$ is
\be
|y_e| = \ft17 \left(  \ft{45}{49} + \ft{9}{49 \sqrt{2}} \right)^{1/3}
\simeq  0.145118 \;.
\ee

Then $a(|y|)$ is well behaved for $0 \leq |y |\leq |y_e|$.
However $a^3$ becomes complex and the scalar curvature ${\cal R}$ 
becomes infinite for some $|y_s| > |y_e|$.
Looking at the explicit analytic expressions for $a$ and ${\cal R}$ 
one sees that this happens, independently  of our concrete choice
of parameters,  because
$\sqrt{H_U^2 - 4 H_S H_T}$ vanishes and then becomes complex. 

 In our concrete numerical example the equation 
$H_U^2 - 4 H_S H_T = 0 $ has two roots, the relevant being
\be
|y_s| = \ft1{23} \left( 8 ( \ft{45}{49} + \ft{9}{49 \sqrt{2}})
\right)^{1/3} - 3 \sqrt{2} \left( \ft{45}{49} + \ft{9}{49 \sqrt{2}}
\right)^{1/3} \simeq 0.165949 \;,
\ee
such that indeed $|y_s| > |y_e|$. 
As we explained above this holds generically for solutions running
into the direction of the enhancement boundary.
%Note that this is not the result of fine-tuning
%parameters. In fact for all choices of parameters satisfying our
%simplifying assumption
%$H_T = 2H_S$ it is manifest that symmetry enhancement happens 
%before $\sqrt{ H_U^2 - 4 H_S H_T} $ vanishes and causes the  
%space-time singularity.
{\it Whenever the solution runs into  
the specific boundary of moduli space, where gauge symmetry is enhanced,
then it reaches this boundary
before the space-time curvature
becomes infinite. This is an example where a moduli space boundary shields a
space-time singularity}. 

The analytical values of $a,{\cal R}$ at $|y_e|$ in the example are:
\be
a(|\tilde y|) = \left(
\ft37 ( \ft{45}{49} + \ft{9}{49 \sqrt{2}})
\right)^{1/6}
\mbox{   and   }
{\cal R}(|\tilde y|) = \ft73 ( \ft73 )^{1/3} \;.
\ee
The analytical expressions of $a,{\cal R}$ at $| y_s|$ for our 
example are complicated  and 
therefore we do not display them.
It is however instructive
to plot various quantities for our specific set of parameters.

\begin{figure}
\centering \scalebox{0.8}{
\includegraphics{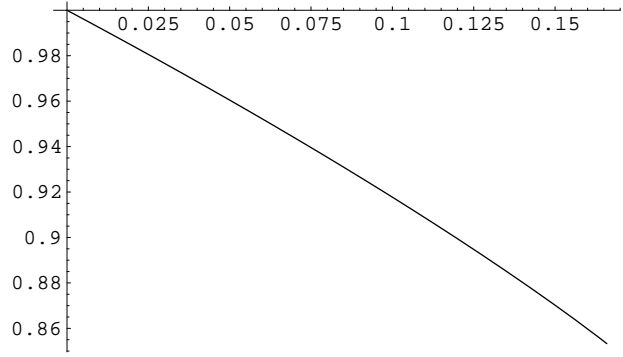}}
\caption{The function $a(y)$ for $0 \leq y \leq 0.17$.
\label{newplot1}}
\end{figure}

\begin{figure}
\centering \scalebox{0.8}{
\includegraphics{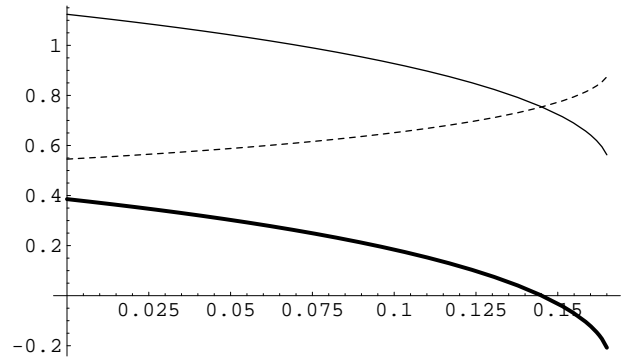}}
\caption{The K\"ahler moduli $h^1(y)$ (dashed line), $h^2(y)$ 
(thick line) and $h^3(y)$ (thin line) for $0 \leq y \leq 0.17$.
At $y = y_c \simeq 0.145118$ the four cycle associated to $h^2$
has collapsed. Note that this happens
before the space-time singularity occurs at $y=y_s \simeq 0.165949$.
%The fact that $h^1(y_e) = h^3(y_e)$ is due to the specific choice of
%parameters that we made in order to have an analytically simple solution.
\label{newplot3}}  
\end{figure}

\begin{figure}
\centering \scalebox{0.8}{
\includegraphics{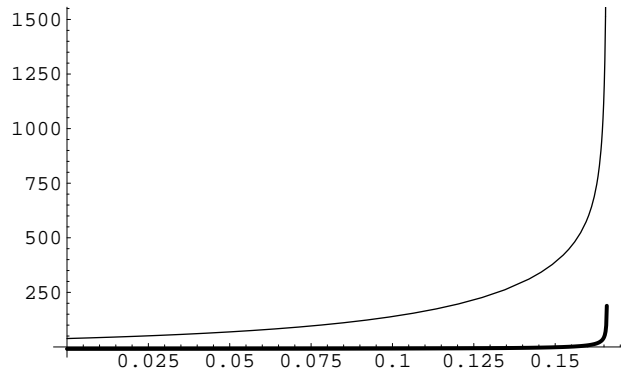}}
\caption{The Ricci scalar $R(y)$ for $0\leq y \leq 0.17$.
The thick line refers to the GST model, the thin line to the HW model.
In both cases the Ricci scalar diverges at $y=y_s \simeq 0.165949$.
\label{newplot4}}
\end{figure}

We display the metric $a$ in Fig.\ref{newplot1}, the moduli $h^1, h^2, h^3$  which solve the generalized attractor equation in Fig.\ref{newplot3} and the space-time curvature ${\cal R}$ in Fig.\ref{newplot4} for
$0 \leq y \leq 0.17$. 
Clearly, the cycle $h^2$ collapses at $|y_c|=|y_e| \simeq  0.145118 \ldots$. 
At this point the space-time is perfectly regular! 
Further down at $|y_s| \simeq 0.165949\ldots$,  
where the cycle $h^2 \simeq   -0.2527\ldots$ is already negative, 
i.e. unphysical, the space-time has a naked singularity. 
{\it All this follows from the solution of the Einstein equation in the 
bulk under condition that we have not yet put the second wall at 
some $|\tilde y|$.}


Note that GST and HW model show the same
qualitative behaviour. This is as expected because the singular
behaviour is due to singularities in the function $a$ and its
derivatives.

{From} the point of view of supergravity nothing special happens
along the line $T=U$ in the scalar manifold. A  negative value of the scalar
field $h^2= T-U$ is as good as a positive one since the metric on the
moduli space at $h^2=0$ is regular and there is no reason to consider 
$T=U$ as a boundary. According to supergravity one can continue
the solution to negative $T-U$ and finally one encounters
a space-time singularity at $|y|=|y_s|$. In order to avoid the singularity
one has to put the second brane at some place $|\tilde y| <| y_s|$, but there
is no distinguished choice of such $|\tilde y|$, nor a physical mechanism
which excises the singularity. Since our solution is supersymmetric, it has zero energy, as shown in \cite{BerKalVPro:2000}, independently of the position of the second brane.

This is different in M/string theory. In M-theory compactified on CY three-folds $h^2$ must be positive as a volume of the cycle. 
$T=U$ is a line of
$SU(2)$ gauge symmetry enhancement and $T-U$ is the associated 
Higgs field. 
Therefore the moduli space ends at $T=U$, negative
values of $T-U$ are related to positive values by the action
of the generator of the Weyl group of $SU(2)$, which is isomorphic
to ${\bf Z}_2$.
This takes particularly  familiar form when using
the dual heterotic description, where $R := \sqrt{T/U}$ is the
radius of the sixth dimension. Therefore  the 
Weyl twist acts as T-duality $R \rightarrow \ft1{R}$
and 
$SU(2)$ gauge symmetry enhancement occurs at the 
selfdual radius, $R = 1$. \footnote{The situation reminds 
some earlier suggestions \cite{BVTV}
to remove the Big Bang singularity using 
$R \rightarrow \ft1{R}$ symmetry of the string theory.}
Since the gauge symmetry enhancement happens at
$y_e < y_s$ it does not make sense to  naively continue to  
$y>y_e$ and in this
way the singularity is excised. Moreover it may be natural to put the
second brane precisely at the enhan\c{c}on point $y=y_e$, defined by the equation 
\be
R(y_e)=1 \ ,\qquad R^2(y)\equiv {T(y)\over U(y)}= {h^1+ h^2\over h^1}
\ee
In this case the ${\bf Z}_2$ orbifold
symmetry acting on $y$ and the Weyl twist / T-duality transformation
acting on the moduli coincide. 
By putting the second wall  at the enhan\c{c}on point $y=y_e$
we enforce the  physics to depend
on $|T-U|$. When putting the second wall at a different place
we would break T-duality spontaneously.

So far we have worked with the prepotential  $STU + \ft13 U^3$ valid inside
the K\"ahler cone. We found that both the resulting theory and the 
domain wall solutions were regular at $T=U$. However we had to stop
there because we reached a boundary and new physics occurred. One way to
capture this new physics is to use the  $T \leftrightarrow U$ symmetric form
of the prepotential that was found in \cite{AntFerTay:1995} 
in the context of heterotic string theory on $K3 \times S^1$:
\be
{\cal V} = STU + \ft13 U^3 \theta( T - U ) + \ft13 T^3 \theta (U - T) \;.
\label{HetPre}
\ee
This is now valid for both positive and negative $T-U$. The build--in
$T \leftrightarrow U$ symmetry reflects that 
negative $T-U$ is related to positive $T-U$ by a large gauge transformation.
The resulting discontinuities are consequences of the $SU(2)$ gauge symmetry
enhancement and reflect the presence of extra massless 
states at $T=U$. They are analogues of the logarithmic branch cuts one 
encounters in four dimensions \cite{AntFerTay:1995}.\footnote{This applies
equally to the $T=U$ line in the $(12,12)$, $(13,11)$ and $(14,10)$ model. 
Note that the new physics at the other boundaries is not taken care of.}

Above we mentioned that it may    be  natural to put the second wall 
at the enhan\c{c}on locus so that $|\tilde y|=|y_e|$. We can use the heterotic prepotential
(\ref{HetPre}) to give an additional argument for this. Namely, the
presence of the discontinuities in the prepotential automatically 
causes a $\delta$-function singularity in the space-time geometry
of a domain wall which tries to cross the boundary $T=U$.
Therefore the enhan\c{c}on itself acts like a source. Note that this 
kink singularity is
different from the naked singularities of the supergravity solution that
we want to excise.

To see this explicitly we first recall that singularities of
the Ricci scalar come from singularities of $a''$, where
$'$ is the derivative with respect to $y$ and 
$
a(y) = ({\cal V}(\tilde{h}(y)))^{1/3} \,.
$
Singularities in $a''$ can therefore descend from 
the $\theta$-function which are present in (\ref{HetPre}) through application
of the chain rule. To work this out we need to be more precise about
how ${\cal V}$ behaves as a function of $\ti{T}-\ti{U}$. Despite the presence
of the $\theta$-functions, ${\cal V}$ itself is actually continuous,
but its derivative with respect to $\ti{T}-\ti{U}$ has a finite jump
at $\ti{T}=\ti{U}$. 
Consequently the second derivative gives a $\delta$-function:
$
\ft{ \der^2 {\cal V}}{\der{(\ti{T}-\ti{U})} \der{(\ti{T} - \ti{U} )}} =
- \left( \ft{\ti{T} + \ti{U}}{2} \right)^2 \delta(\ti{T} - \ti{U})
+ \mbox{finite} \;.
$
This contributes to $a''$:
$
a'' = \ft13 {\cal V}(\ti{h})^{-2/3} {\cal V}'' + \mbox{finite}
\simeq 
\ft{ \der^2 {\cal V}}{\der{(\ti{T}-\ti{U})} \der{(\ti{T} - \ti{U} )}} 
\left[(\ti{T}-\ti{U})' \right]^2 + \mbox{finite}  \;,
$
where we dropped terms, both additive and multiplicative, that stay 
finite for $\ti{T}=\ti{U}$.
Since $\ti{T}-\ti{U}$ has $y=y_e$ as its only zero we find
\be
a'' \sim - \left( \ft{\ti{T} + \ti{U}}{2} \right)^2
(\ti{T} - \ti{U})' \delta(y - y_e) \;.
\ee
This would justify our assertion that it is natural to put the second wall at the enhancement point so that $|\tilde y|= |y_e|$. Any other position   will break the $T$-duality symmetry.

By five-dimensional heterotic -- M-theory duality we expect that the
physics of $SU(2)$ enhancement can be equivalently described in the
M-theory language. In the context of Calabi-Yau compactifications
$SU(2)$ gauge symmetry enhancement (with $g\geq 0$ additional hypermultiplets)
occurs when a divisor collapses into a (genus $g$) curve of $A_1$
singularities. In our case we know from the heterotic analysis
that this curve must have genus 0. The Weyl group ${\bf Z}_2$ is 
encoded in the geometry through the local form of the $A_1$ singularity,
${\bf C}^2 / {\bf Z}_2$. It seems that the Weyl reflections
relating positive  
to negative $T-U$ in the heterotic language
correspond
to the `elementary transformations' discussed in \cite{KatMorPle:1996}.
The extension of the range of moduli as done in (\ref{HetPre})
presumably corresponds to the procedure of gluing in a reflected 
K\"ahler cone at the enhancement boundary, which is described in
\cite{KatMorPle:1996}.

 In this paper we have shown that there is a stringy mechanism which
in certain cases excises space-time singularities which plague
supergravity solutions.
The mechanism is based on the fact that the stringy moduli space
has a boundary on which the moduli space metric is finite. Whereas
this boundary doesn't have a particular meaning in supergravity,
so that solutions can be continued beyond until a space-time
singularity occurs, one encounters new physics at the boundary
in string theory, which makes the space-time singularity  unphysical.

This observation leads to a variety of new issues which have
to be addressed in the future. Most importantly one would like
to understand in detail how the new M/string theory  physics modifies
space-time geometry and excises the singularity. Since $SU(2)$
gauge symmetry enhancement occurs at the boundary, the situation
resembles the enhan\c{c}on  geometry\cite{Johnson:2000qt} and it would be
interesting to explore how far this parallel goes. There are some
further
facts which might be relevant. In particular at the boundary the 
tensionless magnetic strings are present in addition to charged
massless gauge bosons: it was shown in  \cite{critical} that the magnetic string states with charges $\pm( 1, -2, 1)$ have a vanishing tension. Also one should take into
account that the five-dimensional prepotential is purely cubic for  five
non-compact dimensions. However in our domain wall set-up the
fifth dimension is compact and subject to an orbifold projection
which reduces the number of unbroken supersymmetries. Thus the
new stringy physics at the boundary might be more complex and
more interesting than naively expected.

\

We are very grateful to I. Antoniadis, S. Dimopoulos, S. Kachru, 
A. Linde,  E. Silverstein, L. Susskind and N. Toumbas 
for useful discussions. T.M. would like to thank Y. Zunger for 
help in using Mathematica. This work 
was supported by NSF grant PHY-9870115.

\end{document}